\documentclass[prd,eqsecnum,floatfix]{revtex4}
\usepackage{amsfonts}
\usepackage{amssymb}
\usepackage{graphicx}
\newcommand{\be}{\begin{equation}}
\newcommand{\ee}{\end{equation}}
\newcommand{\bear}{\begin{eqnarray}}
\newcommand{\eear}{\end{eqnarray}}

\newcommand{\vev}[1]{\left\langle #1\right\rangle}

\newcommand{\lapproxeq}{\lower .7ex\hbox{$\;\stackrel{\textstyle
<}{\sim}\;$}}
\newcommand{\gapproxeq}{\lower .7ex\hbox{$\;\stackrel{\textstyle
>}{\sim}\;$}}
\newcommand{\stackdown}[2]{\lower 1.4ex\hbox{$\;\stackrel{\textstyle{#1}}
{\scriptstyle{#2}}\;$}}
\newcommand{\beq}{\begin{equation}}
\newcommand{\eeq}{\end{equation}}
\newcommand{\lsp}{\tilde{\chi}}

\newcommand{\bea}{\begin{eqnarray}}
\newcommand{\eea}{\end{eqnarray}}

\makeatletter
\def\slash{\@ifnextchar[{\fmsl@sh}{\fmsl@sh[0mu]}}
\def\fmsl@sh[#1]#2{%
  \mathchoice
    {\@fmsl@sh\displaystyle{#1}{#2}}%
    {\@fmsl@sh\textstyle{#1}{#2}}%
    {\@fmsl@sh\scriptstyle{#1}{#2}}%
    {\@fmsl@sh\scriptscriptstyle{#1}{#2}}}
\def\@fmsl@sh#1#2#3{\m@th\ooalign{$\hfil#1\mkern#2/\hfil$\crcr$#1#3$}}
\makeatother

\begin{document}

\rightline{ACT-09-06, MIFP-06-34}
\rightline{hep-ph/0612152}

\title{Smoothly evolving Supercritical-String Dark Energy
relaxes Supersymmetric-Dark-Matter Constraints}

\author{A.B. Lahanas}
\affiliation{University of Athens, Physics Department,
Nuclear and Particle Physics Section,  GR-157 71,
Athens, Greece.}

\author{N.E. Mavromatos}
\affiliation{King's College London, University of London,
Department of Physics, Strand WC2R 2LS, London, U.K.}

\author{D.V. Nanopoulos}
\affiliation{George P.\ and Cynthia W.\ Mitchell Institute for
Fundamental
Physics, Texas A\&M
University, College Station, TX 77843, USA; \\
Astroparticle Physics Group, Houston
Advanced Research Center (HARC),
Mitchell Campus,
Woodlands, TX~77381, USA; \\
Academy of Athens,
Division of Natural Sciences, 28~Panepistimiou Avenue,
Athens 10679,
Greece}

\begin{abstract}

We show that Supercritical-String-Cosmology (SSC)  off-equilibrium and time-dependent-dilaton effects
lead to a smoothly evolving dark energy for the last 10 billion years in concordance
with all presently available astrophysical data. Such effects dilute by a factor O (10)
the supersymmetric dark matter density (neutralinos), relaxing severe WMAP 1,3 constraints
on the SUSY parameter space. Thus, LHC anticipated searches/discoveries may discriminate
between conventional and Supercritical-String Cosmology.

\end{abstract}


\maketitle


\section{Introduction}

It is not a secret that the Universe is presently expanding in an accelerating mode.
Evidence from the SnIa projects~\cite{snIa} and WMAP 1,3~\cite{wmap} is accumulating in a rather impressive pace,
confirming the existence of dark energy, the fuel that drives acceleration, and shedding
light on its properties.

Fundamental issues, such as the time-(in)dependence of the Dark Energy, 
its equation
of state (w-parameter) and its present value, suspiciously close to 
the critical
or matter density, start to get illuminated.

In a nutshell, recent data indicate that for the last 9 billion years ($z<1.6$) rapidly
evolving dark energy is ruled out, while negative pressure ($w<0$), the hallmark
of dark energy appears to be present.

It is a common secret, at least among theorists, that such properties,
as indicated above, of Dark Energy is not exactly a bowl full of cherries.
Concentrating on critical string theory or, its up-grading, M-Theory, that
provides a rather solid framework for Quantum Gravity and Cosmology, it is
well-known that a "Cosmological Constant" or even an expanding Robertson-Walker-Friedman (RWF) Universe
is not easily admissible.

The basic steps for a correct formulation for an expanding RWF Universe
in string theory were taken in~\cite{aben}, where the crucial role of a
time-dependent dilaton field, actually linear in time, has been emphasised,  thus
establishing Supercritical String dilaton Cosmology (SSC)~\cite{emninfl,gasperini}. That led naturally
to dissipative Liouville-String Cosmology or Q-Cosmology~\cite{diamandis,diamandis2,emnw}, involving super (non)
critical string~\cite{aben} cosmological backgrounds, with the identification~\cite{emn} of target time with the
world-sheet zero mode of the Liouville field~\cite{ddk}. Such cosmologies were found
to asymptote (in cosmic time) the conformal backgrounds of \cite{aben},
which are thus viewed as equilibrium (relaxation) configurations of
the non-equilibrium cosmologies~\cite{diamandis,emnw}. This implies relaxation of the associated Dark Energy of
such cosmologies and a gravitational friction associated with conformal theory central
charge deficit Q. For a recent review we refer the reader to~\cite{emnw} where concepts and methods are outlined
in some detail.

In this letter, we concentrate on the Supercritical-String Cosmological (SSC)
off-equilibrium and/or time-dependent dilaton
effects on the Boltzman equation describing relic abundances
and the associated particle-physics phenomenology~\cite{lmnbolt}. We show
first that off-equilibrium supercritical string cosmologies,
which notably are consistent with the current astrophysical data
from supernovae, as demonstrated recently~\cite{emmn}, predict a rather smoothly
evolving Dark Energy at least for the last ten billion years ( $0<z<1.6$ ), in accordance with the
very recent observations on supernovae~\cite{riessnew}. We then proceed to show that
recent WMAP 1,3~\cite{wmap} severe constraints on the available phase-space distribution
of favourite supersymmetric Dark Matter candidates~\cite{susyconstr}, namely the neutralinos~\cite{EHNOS} are relaxed when off-equilibrium effects, are considered. Actually,
we find that the neutralino (Dark Matter) density is diluted by a factor O (10), while
the baryon density is diluted by a factor O (1) (!). 
Thus, neutralino Dark matter enhances its odds and LHC has
more available SUSY parameter to exploit.

\section{Solving the Super-Critical String Cosmological equations}

The dynamical equations in the framework of the Supercritical String Cosmology (SSC), and their derivation, have been given in a previous publication~\cite{diamandis2}, where we refer the interested reader for details.
We repeat here only the final result for the sake of completeness.
By combining  the available non-critical string equations we can cast them in the form of a first
order dynamical system taking as variables $\hat{H},\phi,\dot{\phi},Q,{\tilde{\varrho}}_m $.
$Q$ is the central charge deficit and ${\tilde{\varrho}}_m$ the densities of all species involved with the exception of the dilaton energy density which has been already counted for:
\bear
&&\ddot{\phi}= -2 \hat{H}^2-3 \hat{H} \dot{\phi} - e^{\phi} Q ( \dot{\phi}+\hat{H})+
\frac{1}{2} ({\tilde{\varrho}}_m +{\tilde{p}}_m ) \nonumber  \\
&& 3 \dot{\hat{H}}=-\hat{H}^2-2 {\dot{\phi}}^2+e^{\phi} Q ( \dot{\phi}+\hat{H})
- \frac{1}{2} (3 {\tilde{\varrho}}_m +{\tilde{p}}_m ) \nonumber \\
&&{ \dot{\tilde{\varrho}}_m }+
2 Q \dot{Q} e^{2 \phi} =
 -3 \hat{H} ( {\tilde{\varrho}}_m +{\tilde p}_m)
+\dot{\phi}\;({\tilde{\varrho}}_m - 3 {\tilde{p}}_m )+ \nonumber \\
&&4 \;(\hat{H}+\dot{\phi})\;
( -\hat{H}^2+ {\dot \phi}^2 + e^{\phi} Q ( \dot{\phi}+\hat{H}) + {\tilde{p}}_m \; )\nonumber  \\
&&(\;e^\phi Q + \hat{H}\;) e^\phi \;\dot{Q}\;=\; \nonumber \\
&&\frac{{ \dot{\tilde{p}}_m }}{2}+
\frac{1}{2} (e^\phi Q - 2 \hat{H})\;{ {\tilde{\varrho}}_m } +
\frac{1}{6} (e^\phi Q + 22 \hat{H} +18 \dot{\phi}) \;{ {\tilde{p}}_m } + \nonumber \\
&& [\;
-\frac{e^{2 \phi}}{3} (\dot{\phi}+\hat{H})\; Q^2 + e^{\phi}\; (\frac{23}{3} \; {\dot{\phi}}^2
+ 8 \hat{H}^2 + \frac{47}{3} \;\dot{\phi} \hat{H}) \;Q + \nonumber \nonumber \\
&& 10 \;{\dot{\phi}}^3 + \frac{62}{3}\; {\dot{\phi}}^2 \hat{H}
+12\; {\dot{\phi}} \hat{H}^2 + \frac{4}{3} \;\hat{H}^3 \;] \quad . \label{diffs}
\eear
The above equations have been expressed in terms of the dimensionless Einstein time~\cite{aben} $t_E$ which is related to the cosmic time $t$ of the RWF Universe through~\cite{diamandis2} $t_E \equiv \omega \; t$. $\omega $ is in principle arbitrary and it has dimensions of inverse time. The hatted density and pressure appearing above are actually the corresponding quantities multiplied by $8 \pi G_N / \omega^2$. A convenient choice is $\omega = \sqrt{3} H_0$ where $H_0$ is the Hubble constant which by defining the  rescaled Hubble constant $h_0$ can be written as $H_0=1.022 \times 10^{-10} h_0 \; yr^{-1}$. With this choice for $\omega$ the hatted quantities  are scaled with the critical density $\rho_c=3 H_0^2 \; /\; 8 \pi G_N$. 
In this system of units one year of cosmic time corresponds to $t_E=1.292 \times 10^{-10}$ and one second to $t_E=4.097 \times 10^{-18}$. Thus in one unit of the Einstein time we can encompass almost the whole Universe history ! In practice this means that any effort to get reliable numerical results should be based on a very fine coarse-graining of $t_E$. Viewed from this point the use of the redshift as an  independent variable is more efficient for the numerical manipulations. One has merely to convert
the derivatives with respect $t_E$ to those with respect the redshift $z$ in order to have the equations in terms of the redshift.
The Hubble function in the Einstein frame $\hat{H}$ is defined by $\hat{H}={d ln \;a}/{dt_E}$,  and is  related to the Hubble expansion rate of the ordinary Cosmology $H$ through $\hat{H}=H/\omega$. With the value for $\omega$ quoted above its today value is  $\hat{H} |_{today}= 1/\sqrt{3}$. The equation relating the Hubble expansion rate to the densities of the various species involved receive, in this system of units,  the following form,
\bear
3 \; {\hat{H}}^2 \;=\; \Omega +\Omega_{\phi}+\Omega_{n.c.} \; , \label{hh}
\eear
which is not an independent equation since it follows by combining the equations \ref{diffs}.
In Eq. (\ref{hh}), $\Omega$ is the sum of the ratios  of the densities of the radiation and any sort of matter to the critical density $\rho_c$, while $\Omega_{\phi}, \Omega_{n.c.} $ are the corresponding quantities for the dilaton field and the non-critical terms respectively. The latter are certainly absent in the critical string theory. 
We recall that the critical density is $\rho_c = 8.097 \times 10^{-11} \;h_0^2 \; eV^4$. $\Omega$ includes the density of the radiation given by $\Omega_r \; h_0^2= 4.063 \times 10^{45} \; g_{eff} {\left( T/GeV \right) }^4$. 
For temperatures less than $1 \; MeV$ only photons and neutrinos contribute and $g_{eff}=2.91$. 
Today, $T_{CMB}=2.725 \;^0K=2.349 \times 10^{-13} \; GeV$, and $\Omega_r \; h_0^2= 3.59 \times 10^{-5}$, of which $2.47 \times 10^{-5}$ is carried by photons and $1.12 \times 10^{-5}$ by neutrinos. 
In addition to the radiation, $\Omega$ includes the contributions of the non-relativistic matter and perhaps other forms of matter which we call "exotic" and we assume that they directly feel the effect of the non-critical terms. 
Every species is subject to a different continuity equation derived from  
the third of the equations (\ref{diffs}), which is actually the continuity equation, and it includes all species involved within the density ${\tilde{\rho}}_m$. To be more specific, 
assuming an equation of state for the matter $w=0$, radiation $w=1/3$ and "exotic" matter~\cite{emnw,emmn,diamandis2}, which is the remaining part in the energy density that directly feels the effect of the non-critical terms with unknown $w$ parameter, one has three equations. This covers the more general situation. One can of course abandon the idea of the exotic matter by adding matter and exotic matter and attributing $w=0$ to the exotic piece as well~\cite{diamandis2}. 

One should note that the continuity equation for the radiation density $\rho_r$ depends neither on the dilaton nor on the non-critical terms and retains the form of the traditional no-dilaton cosmology which entails to $\rho_r \; a^4 = \mathrm{constant}$. However matter, $w=0$, does feel the effect of the dilaton dynamics. In fact  matter $\rho_M$ and radiation $\rho_r$ satisfy the following continuity equations 
\bear
&&\dot{\rho}_M
 +3 \hat{H} {\rho}_M
-\dot{\phi}\;{\rho}_M  \;=\;0 \nonumber \\
&&\dot{\rho_r}+4 \hat{H} {\rho}_r \;=\;0
\eear
This holds even in the absence of the non-critical terms, as is the case of critical-strings dilaton cosmologies~\cite{gasperini,diamandis2,lmnbolt}. This can be seen from the third of the equations (\ref{diffs}) if the last term in it, which is the contribution of the non-critical terms, is omitted and the charge $Q$ is set to a constant. 
The exotic matter is obeying an equation of state which is read from the same equation, assuming an equation of state $p_e=w_e \; \rho_e$ relating its pressure $p_e$ to its density $\rho_e$, and it is affected by both dilaton and the non-critical terms. For lack of space we do not present it here.

\begin{figure}[H]
\begin{center}
\includegraphics[width=14cm]{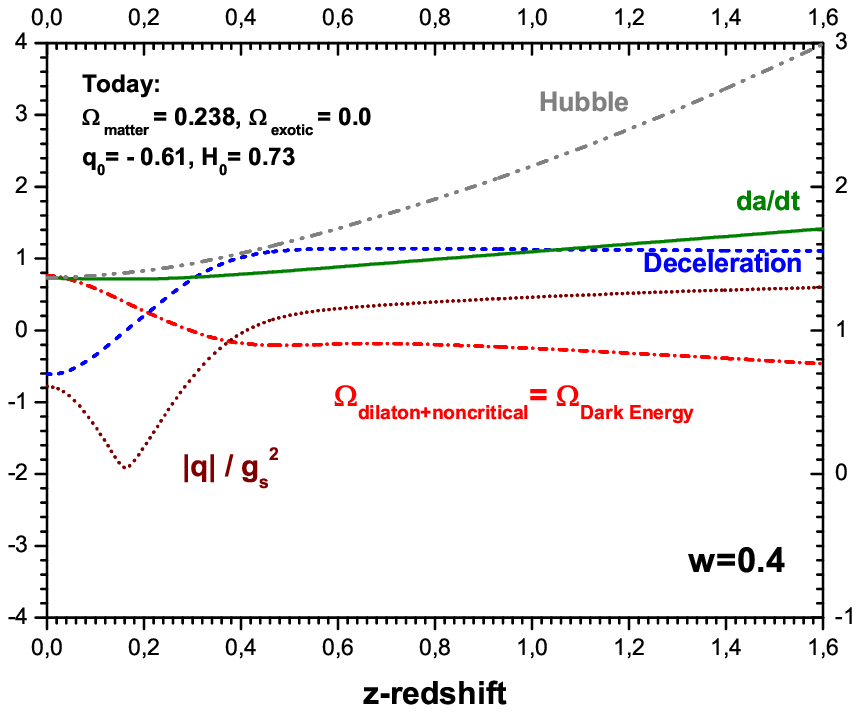}
\end{center}
\caption[]{
The energy density carried by the dilaton and the non-critical terms ( dashed-dotted red  line), the deceleration  (~dashed blue line), the rescaled Hubble expansion rate ( dashed - double dotted grey line~) 
and the derivative of the cosmic scale factor ( solid green line ) as functions of the redshift in the range $0 < z < 1.6$ are displayed. Their values refer to the left y-axis. The ratio $|q|/g_s^2$ is also displayed  ( dotted  brown line ) with values on the right vertical axis.
}
\label{fig0}
\end{figure}

Before embarking on discussing the details of our computation we should remark that the parameter 
$w \equiv w_e$ of exotic matter is the only free parameter in this consideration and any choice for it leads to a different prediction which oughts to comply with the current  astrophysical observations. Of special importance are the current data from supernovae observations~\cite{riessnew} which predict a rather smooth evolution of the Dark Energy, at least in the regime $0 <z < 1.6$, and cosmological models based on String Theory should agree with this observation. This information comes as an additional constraint, the other being the behaviour of the acceleration of the Universe which according to the existing data is started at redshifts $z \sim 0.20$. As a sneak preview, in figure \ref{fig0} we display the behaviour of the Dark Energy (DE) for redshift values  $0 < z < 1.6$ assuming that DE is carried by the dilaton and the non-critical terms. It is remarkable that such a smooth evolution follows only for values of $w$ in the vicinity of $w \simeq 0.4$, which we assume in the following, and therefore $w$ is fixed by these data. 
This could not have been foreseen. What is equally important,  
as we will discuss later, values of $w$ in this range affect the supersymmetric prediction for Dark Matter (DM), due to the modification of the Boltzmann equation from the dilaton and the non-critical terms , with important consequences for the cosmological and phenomenological implication of supersymmetric models. 
For completeness in the same figure we display the Hubble expansion rate $H(z)$ and the deceleration $q(z)$ as well as the variation with time, $da/dt $, of the cosmic scale factor. We also display the ratio $| q |/ g_s^2 $ which as at late eras yields a precise relation of the deceleration $q$ with the string coupling constant $g_s$ \cite{emnw,emninfl}.

 With the above in mind, we can now proceed towards solving the system of the pertinent differential equations to find the evolution of the Super-Critical String Universe. The details have been given in \cite{diamandis2} and will not be repeated. In that work the solution has been expressed in terms of the
Einstein time. Here instead we solve the equations in terms of the redshift, rather than the Einstein time, which is a much better parameter to use  for the numerical treatment as stated earlier.
After doing this, we shall discuss  the compatibility of the non-critical string predictions with the cosmological data, as well as  the modification of the predictions for the CDM ( Cold Dark Matter ) abundances, completing the analysis of \cite{lmnbolt}.
The equations in terms of the Einstein variable $t_E$ can be converted to those in which the relevant independent parameter is the redshift, as we advertised, through the redshift-time relation
\bear
t_E \;=\; \int_z^{\infty} \frac{dz}{(1+z)\;\hat{H}} \; \;\;. \label{z2te}
\eear
which in differential form is
\bear
\frac{t_E}{dz} \;=\;- \frac{1}{(1+z)\;\hat{H}} \; \;\;. \label{z2tex}
\eear
The latter can be inverted and integrated to yield the redshift as a function of the Einstein time (and hence the cosmic time), if so wished.
As is well known, the relation of the redshift $z$ to the cosmic scale factor $a$ is given by
\bear
z+1\;=\;\frac{a_0}{a} \; . \label{redit}
\eear
If we want to relate it to the thermal history of the Universe we recall that 
the total energy density of the radiation is given by~\cite{kolb}
\bear
\rho_r\;=\;\frac{\pi^2}{30} \;g_{eff}(T) \;T^4  \label{geff}
\eear
where $g_{eff}$ counts the relativistic degrees of freedom and $T$ is the temperature of the ``photon gas". This is measured by antennas and satellites and its value today is accurately known, $T_{CMB}=2.725 \; ^0K$. Since $a T $ and $\rho_r a^4 $ remain constants, the redshift-temperature relation is given by
\bear
z+1\;=\; \; {\left( \frac{g_{eff}(T)}{g_{eff}(T_{CMB})} \right)}^{1/4} \; \frac{T}{T_{CMB}}  \; . \label{redit2}
\eear
With the central value for $T_{CMB}$ quoted above,
$T/T_{CMB}=0.4257 \times 10^{13} \;(T/GeV)$ if the temperature is given in $GeV$.
For temperatures lower than the neutrino decoupling temperature, $T_d \approx 2 \; MeV$, we have
\bear
g_{eff}=2+\frac{7}{4} \; N_{\nu} \left( \frac{T_{\nu}}{T} \right)\;
\eear
where $T_{\nu}$ is the neutrino temperature.
Using the fact that $T_{\nu}/T= {\left( 4/11\right)}^{1/3}$, in this temperature region, the value for $g_{eff}$ above is $g_{eff}=2.91$, for two massless neutrino and antineutrino species ($N_{\nu}=2$) while it equals to $g_{eff}=3.37$ if the third generation of neutrinos and antineutrinos is also considered massless ($N_{\nu}=3$). On the other hand for high temperatures, well above the typical supersymmetry breaking scale $M_{S}$, $g_{eff}\approx230$ provided that we are below the GUT scale and the particle content is that of the minimal supersymmetric model. Therefore for such high temperatures ${g_{eff}(T)}/{g_{eff}(T_{CMB})}=79$ and the redshift-temperature in this regime, $M_{GUT}>T>M_{S}$,  is
\bear
z+1\;=\; \; 1.27 \times 10^{13} \; \frac{T}{GeV}   . \label{redit3}
\eear
with $T$ given in $GeV$. Eq. (\ref{redit3}) merely states that to reach temperature as large as $10^6 \; GeV$ one needs explore the solutions of the cosmological equations for redshift values approaching $z \sim 10^{19}$.

The initial inputs in solving  the  system of the first order differential equations are the densities of the radiation, or equivalently the cosmic microwave background temperature
$T_{CMB}$ on which this depends on as described previously, the  matter and the exotic matter densities today ($z=0$), as well as the rescaled  Hubble constant $h_0$ and the deceleration $q_0$. Additional inputs are the initial dilaton value $\phi_0$ and the value of $w$ governing the equation of state of the exotic matter. Regarding the initial value of the central charge deficit $Q_0$, we remind the reader that~\cite{diamandis,diamandis2} it is not an independent input, but is determined in terms of the $q_0, h_0, \phi_0$ through a quadratic algebraic equation, which actually holds at all times, and is derived from the set of differential equations at hand. Its form is given by
\bear 
2 \;{{Q}}^2 - e^{- \phi} \hat{H} \;{Q} + e^{- 2 \phi}\; ( \;{\dot{\phi}}^2 -
8 {\hat{H}}^2 - 3 \hat{H} \dot{\phi}+
\frac{5}{2} {\tilde{\varrho}}_m + \frac{1}{2} {\tilde{p}}_m \;) \;=\;0 
\label{quadratic} 
\eear
and it actually follows by combining all available equations. This along with the equation giving the deceleration in terms of the other parameters involved, 
\bear 
q\;=\; - \frac{1}{{\hat{H}}^2} \; \left(\; \frac{2}{3}\; {\hat{H}}^2 - \frac{2}{3}\; {\dot{\phi}}^2
+ \frac{e^\phi Q}{3} \; ( \hat{H} + \dot{\phi} ) 
- \frac{1}{2} {\tilde{\varrho}}_m - \frac{1}{6} {\tilde{p}}_m \right) \; ,
\label{decelera} 
\eear
are solved 
to yield the charge $Q_0$, as well as the derivative ${\dot{\phi}}_0$, in terms of $q_0, h_0, \phi_0$ and the densities. Therefore the initial values of the first order differential system (\ref{diffs}) are completely known (for more details see \cite{diamandis2}). 

The non-critical terms, as well as the dilaton density, although non-thermal~\cite{lmnbolt} do gravitate and are related to the Hubble expansion rate through Eq. (\ref{hh}). If the total density is denoted by $\rho_{total}$ then it is convenient to define ${\tilde{g}}_{eff}$ by~\cite{lmnbolt}
\bear
\rho_{total} \equiv \frac{\pi^2}{30} \; {\tilde{g}}_{eff} \;T^4 \;\; .
\eear
Since $3 {\hat{H}}^2 = \Omega_{total}$, by Eq. \ref{hh}, the ratio ${\tilde{g}}_{eff}/ g_{eff}$ is provided by
\bear
\frac{{\tilde{g}}_{eff}}{ g_{eff}} = \frac{3 {\hat{H}}^2}{\Omega_r} \; \; . \label{gtilde}
\eear
The significance of this ratio is that it explicitly appears in the formula determining the freeze-out temperature and the relic densities as has been shown in \cite{lmnbolt}. The solution of the string cosmological equations determine the Hubble expansion rate and the densities of all species involved and hence this ratio can be extracted at any time, or redshift.
Another important quantity which significantly affects the  relic density is a "reduction"
factor~{\footnote{We call it reduction factor, as this will be our case discussed here,  but in general it can also be an ``increase'' factor if it happens to be larger than unity.}} given by~\cite{lmnbolt}
\bear
R\;=\; {\left( \frac{{\tilde{g}}_{eff}^{*}}{ g_{eff}^{*}} \right)}^{1/2} \;
exp \; \left( \; \int_{x_0}^{x_f} \frac{\Gamma {\hat{H}}^{-1}}{x} \; dx \right) 
\; \; . \label{red}
\eear
 The variable $x$, as usual, is the ratio $x=T/m_{\lsp}$ and all quantities in the integral in Eq. (\ref{red}) are considered as functions of $x$. 
$x_f$ is the corresponding freeze-out point and $x_0$ corresponds to 
the present-day temperature of the Universe. Stars denote quantities at the freeze-out point.
In Eq. (\ref{red}), $\Gamma$ is a quantity which equals to $\dot{\phi}$, if the species whose relic refers to is not coupled to the non-critical terms, as we assume for matter $(w=0)$. However it receives additional contributions from the non-critical terms in the case of the exotic matter contributions.
The general form of $\Gamma$ is~\cite{lmnbolt}: 
$\Gamma (t) \equiv  \dot \phi  - \frac{e^{-\phi}}{2}
\left( g^{\mu\nu}{\tilde \beta}_{\mu\nu}^{\rm Grav}
+ 2e^{2 \phi} {\tilde \beta}^\phi \right)$
where the beta function terms are given by
$g^{\mu \nu} {\tilde \beta}_{\mu \nu} + 2 e^{2 \phi}\; {\tilde \beta}^\phi
=-2 \;( \;4\ddot{\phi} + 2 {\dot \phi}^2 + 9 \hat{H} \dot{\phi} +
3 (1-q) {\hat{H}}^2+Q e^{\phi} (4 \dot{\phi}+3 \hat{H}))$.

The role of $R$ is very significant since the relic density of a particular species in the string model under consideration follows by the one obtained in ordinary cosmology multiplied by $R$. In fact we have shown in \cite{lmnbolt} that 
\bear
\Omega_{\tilde{\chi}} h_0^2 \;=\; R \times \left( \Omega_{\tilde{\chi}} h_0^2 \right)_{(0)} 
\; \; .
\label{reduc}
\eear
The quantity on the right, labelled by $(0)$, is that derived in ordinary treatments usually approximated by
\bear
\left( \Omega_{\tilde{\chi}} h_0^2 \right)_{(0)}\;=\;\
\frac{1.066\; \times  10^9\; {GeV}^{-1}}{ M_P \;\sqrt{g^{*}_{eff}} \;J} \; \; ,
\eear
where $J\equiv \int_0^{x_f} \vev{ v \sigma} dx$.
The freeze-out point $x_f$ in the string model which appear in the equations above turns out to be not very different from the one obtained in the standard cosmology. In fact $x_f$ is determined by a modified equation~\cite{lmnbolt}
\bear
x_f^{-1}\;=\; ln \left[ 0.03824 \; g_s\; \frac{M_P m_{\lsp}}{\sqrt{g^{*}_{eff}}}
x_f^{1/2} {\vev{ v \sigma}}_f \right]\;+\;
\frac{1}{2} \;  \left(  ln  \frac{g_{eff}^{*}}{ {\tilde{g}}_{eff}^{*}  } \right) \;+\;
\int_{x_f}^{x_{in}} \;  \frac{\Gamma H^{-1}}{x} \;dx  \quad .  \label{xf}
\eear
in which the ratio given by Eq. (\ref{gtilde}) at the freeze point $x_f$, and the integral
$\int_{x_f}^{x_{in}} \frac{\Gamma {\hat{H}}^{-1}}{x} \; dx$ explicitly
appear. In this integral the upper end of the integration refers to a temperature the Universe had just after inflation. These two terms result to a shift in $x_f$ from the one obtained in the ordinary cosmology.  For ordinary matter as well as Dark Matter, which do not directly couple to the noncritical terms, $\Gamma =  \dot{\phi}$, and for values of the $w$ of the exotic matter close to $w=0.4$, which is cosmologically preferred for reasons explained earlier, the contribution of the last two terms in (\ref{xf})  little affect the freeze out point $x_f$. Actually we found that the contribution of these terms in the r.h.s. of this equation is less than $10 \%$ of the total contribution and almost insensitive to the MSSM inputs. This holds for both hadronic matter and Dark Matter which means that their corresponding freeze-out temperatures are shifted by small amounts. The effect of the reduction factor $R$ is however  more drastic and different in the DM and hadron cases. As is evident from Eq. (\ref{red}) $R$ is affected by two terms which contribute in opposite directions, as we find in our numerical treatment. The exponential factor tends to decrease but this  is countered balanced by the prefactor 
$ {({{\tilde{g}}_{eff}^{*}}/{ g_{eff}^{*}})}^{1/2}$ which tends to increase it. For a typical MSSM case and  having the neutralino as the LSP Dark Matter candidate the exponent yields $\approx 0.015$ while the prefactor $\approx 6.0$ resulting to $R \approx 0.1$. For a hadron with typical mass of $1 \; GeV$ the corresponding numbers are $\approx 0.04$ and $\approx 15.0$ respectively, due to the fact that the freeze-out temperature in the hadron case is smaller,  resulting to a net effect $R \approx 1/2$. 
This conclusion is very important since it shows that in the context of these string models supersymmetry can survive the otherwise stringent constraints imposed by the cosmological data concerning the relic abundance of Dark Matter. Regions of the parameter space in which
the relic density of the CDM in ordinary cosmology is of order unity is not excluded now since the reduction factor renders it to ${\cal{O}}(0.1)$. At the same time the same reduction mechanism does not affect the relics of the ordinary matter leaving the predictions for conventional matter relic abundances unaffected.

In order to quantify our findings we solve Eqs. (\ref{diffs}) taking as independent variable the redshift rather $t_E$ as described earlier.  The initial conditions today, $z=0$, are as follows. For the rescaled Hubble constant and deceleration we take the central values, $h_0=0.73, q_0=-0.61$ while the Universe temperature today is taken $T_{CMB}=2.725^0 K$. Moving within the experimental limits for these quantities causes little change. As stated earlier $T_{CMB}$ yields the photon and neutrino energy densities which contribute to $\Omega$ of Eq. (\ref{hh}). The remaining contribution to $\Omega$ is the matter $\Omega_M$, characterised by an equation of state constant $w=0$ which include both baryonic density and Dark Matter, and the "exotic" matter contribution $\Omega_e$ whose $w \equiv w_e$ is in principle arbitrary allowing to vary within the model as discussed earlier. 
We take as initial conditions $\Omega_M =0.238$ and $\Omega_e=0$ and for these inputs we get solutions which are in agreement with the cosmological data provided $w_e$ is close to $w_e=0.4$. The reduction factor associated with the relic density is of the order of ${\cal{O}}(0.1)$ which has dramatic consequences for the relic density predictions in the Minimal Supersymmetric Standard Model
(MSSM). For the dilaton we can take $\phi_0=0$, without loss of generality, since any non-zero initial value leads to the same solutions with only a rescaled central charge deficit.

\begin{figure}[]
\begin{center}
\includegraphics[width=14cm]{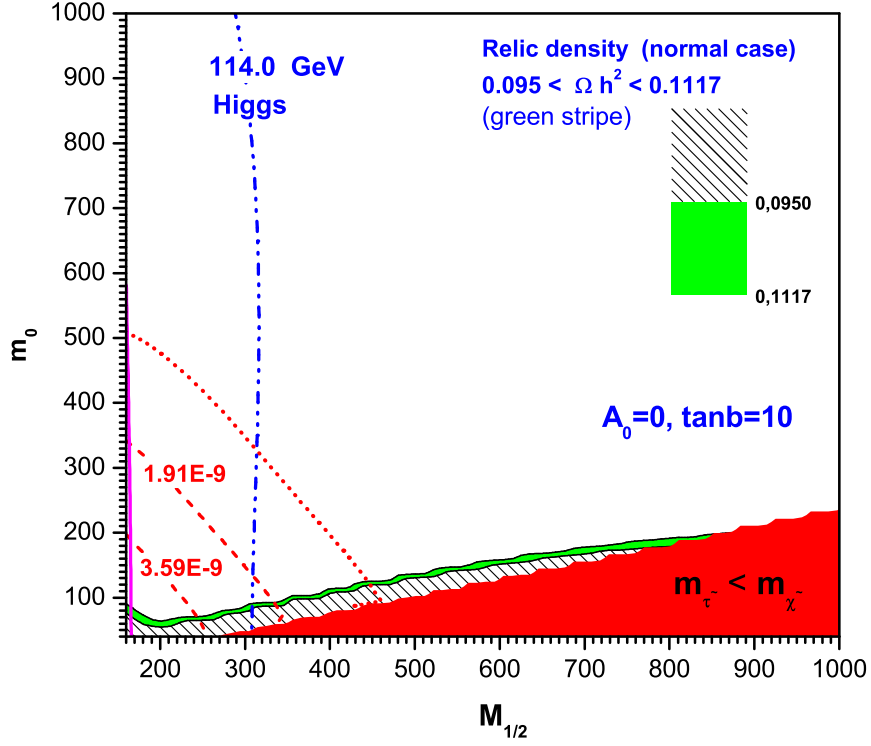}
\end{center}
\caption[]{
In the thin green (grey) stripe the neutralino  relic density is within the  WMAP3 limits
$0.0950 < \Omega_{CDM}h^2 < 0.1117$, for values $A_0=0$ and $tan \beta=10$, according to the conventional calculation. The dashed double dotted line (in blue) delineates the boundary along which the Higgs mass is equal to $114.0 \; GeV$. The dashed lines (in red) are the $1\sigma$ boundaries for the allowed region by the $g-2$ muon's data as shown in the figure. The dotted lines (in red) delineate the same boundaries at the $2 \sigma$'s level. In the hatched region $0.0950 > \Omega_{CDM}h^2$, while in the dark (red) region at the bottom the LSP is a stau.
}
\label{fig1}
\end{figure}

\begin{figure}[]
\begin{center}
\includegraphics[width=14cm]{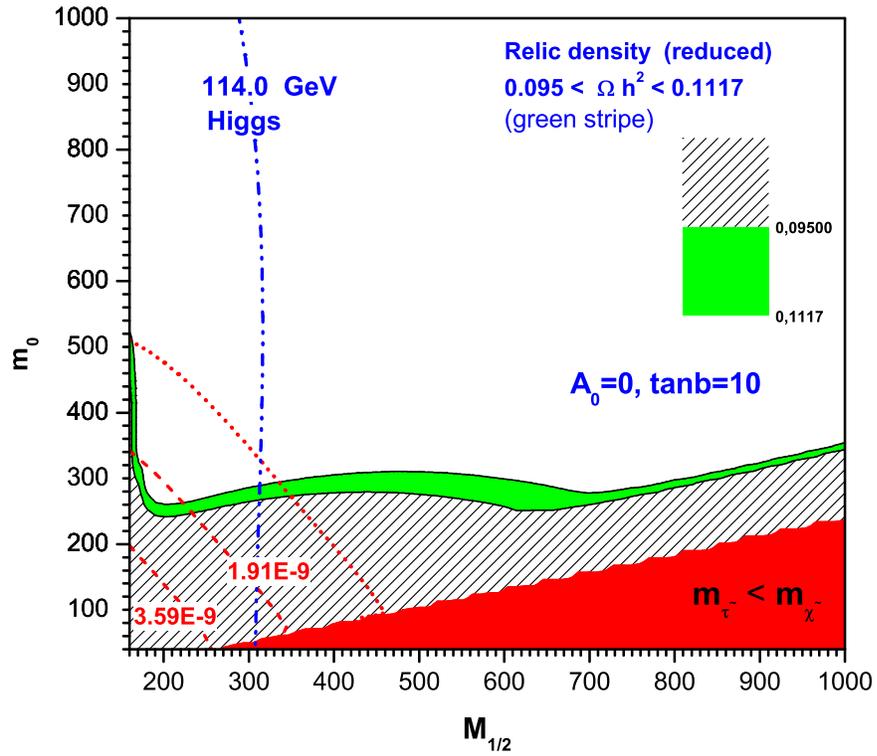}
\end{center}
\caption[]{
The same as in figure \ref{fig1} according to the new calculation in which the relic density is reduced
as described in the main text.
}
\label{fig2}
\end{figure}

\begin{figure}[]
\begin{center}
\includegraphics[width=14cm]{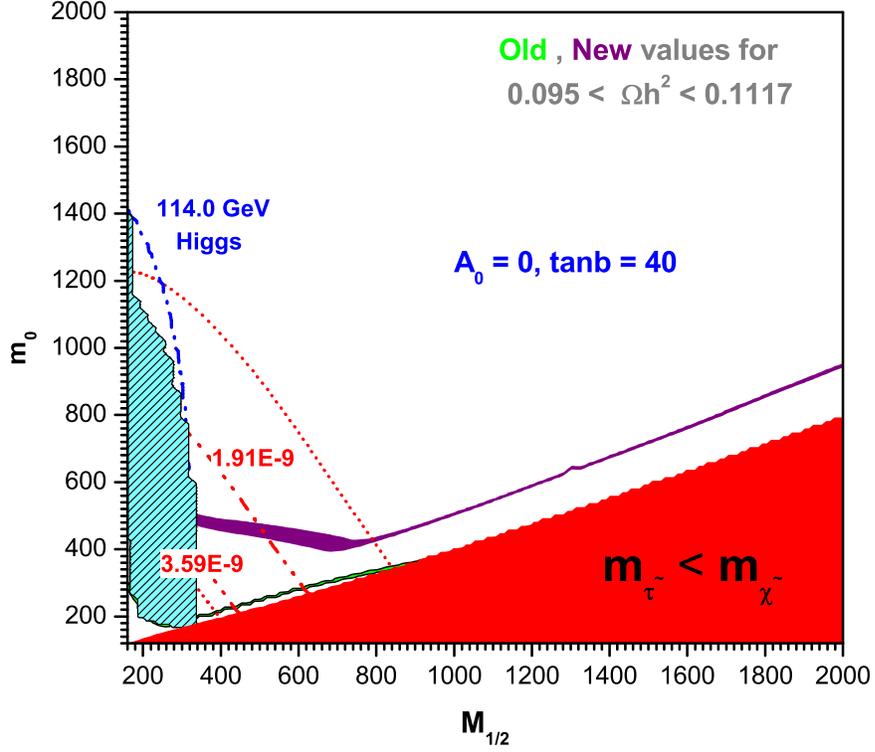}
\end{center}
\caption[]{
In the very thin green (grey) stripe the neutralino  relic density is within the  WMAP3 limits
$0.0950 < \Omega_{CDM}h^2 < 0.1117$, for values of $A_0=0$ and $tan \beta=40$ shown in the figure, according to the conventional calculation. The thin dark (purple) region lying above is the same region according to the new calculation with the reduction factor for the MSSM inputs shown in the figure. The remaining Higgs and $g-2$ boundaries are as in figure \ref{fig1}. The hatched dark (cyan) region on the left is excluded by $b \rightarrow s \; \gamma$ data.
}
\label{fig3}
\end{figure}

\begin{figure}[]
\begin{center}
\includegraphics[width=14cm]{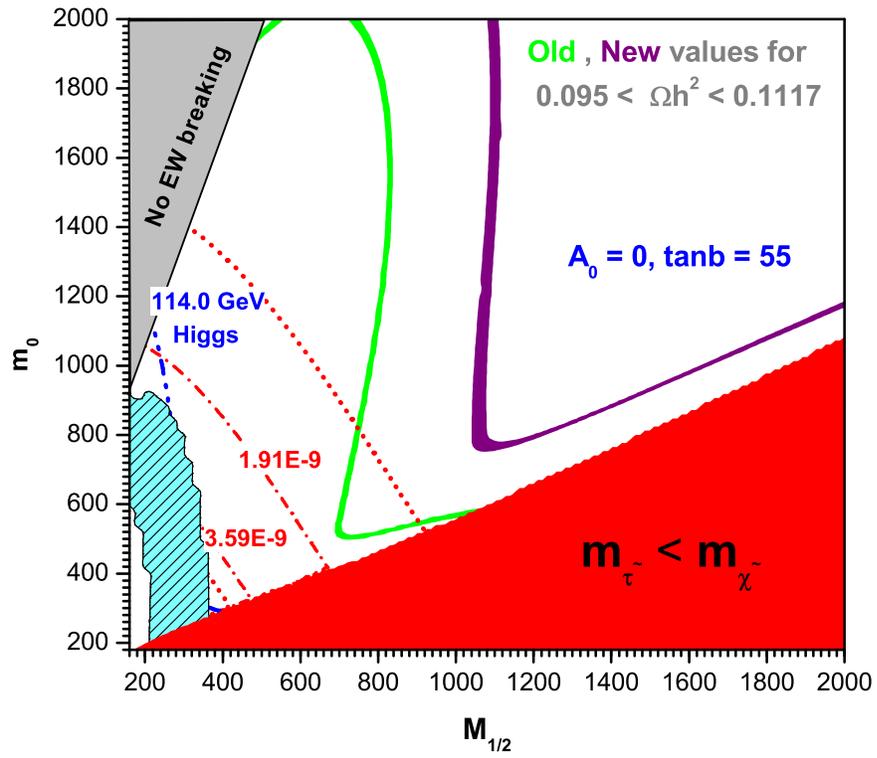}
\end{center}
\caption[]
{
The same as in figure \ref{fig3} for $A_0=0$ and $tan \beta=55$.
}
\label{fig4}
\end{figure}

In figures \ref{fig1} to \ref{fig4} we present representative outputs 
in the $m_0 \;-\; M_{1/2}$ plane for fixed values of the $A_0$ and $tan \beta$ and $\mu>0$. We assume the constrained MSSM with universal boundary conditions at the unification scale. The predictions of this model in conjunction with the DM cosmological constraints has been the subject of numerous studies in the past~\cite{bbik,coannihil,Baer:2002fv,tril,funnels,higgspole,focusit,Baer:2006ff}.

In figure \ref{fig1}, and for values $A_0=0, \; tan \beta=10$, the thin dark stripe (green) is the cosmologically allowed region by the WMAP3 analysis $0.095 < \Omega_{LSP} < 0.1117 $ , as it is calculated in the conventional approach. We also display the allowed region by measurements of the muon's anomalous magnetic moment constraint  $1.91 \times 10^{-9} < \Delta \alpha_{\mu} <  3.59 \times 10^{-9}$ at the $1 \sigma$ level \cite{eidel}. 
The boundaries of this region are shown as dashed lines (in red) on the left lower corner of the panel. The small dotted line ( in red ) marks the boundary of the region for which the lower bound has moved to its $2 \sigma$ limit. The boundary of the region $m_{Higgs} \geq 114.0 \;GeV$ is also shown as a  dashed double dotted line (in blue). The allowed by the Higgs bounds limits stand on the right of this line. In the hatched region below the cosmologically allowed stripe the relic density gets values lower than $0.095$. The $M_{1/2}$ axis has been cut at $160 \; GeV$ by the lower experimental bound put by chargino searches.

The outputs presented in figure \ref{fig1} should be contrasted with those displayed in figure \ref{fig2} where we see the same limits assuming the new approach according to which the relic density suffers a reduction. We clearly see that the new allowed region is now shifted upwards and gets broadened moving therefore the cosmologically interesting region to higher $m_o$ values and away from the coannihilation region. This is due to the fact the reduction factor discussed previously stays almost constant, all over the $m_0 \;-\; M_{1/2}$ plane, getting values around $0.10$. This behaviour is independent of the value $x_{in}$ appearing in Eq. ( \ref{xf}) determining the freeze point, provided that $x_{in}$ corresponds to redshift values  larger  than $z \sim 10^{15}$ . We recall  that the freeze-out point, given by Eq (\ref{xf}), is shifted by only $10 \%$.

In figure \ref{fig3} we present the same results for a different value for $tan \beta=40$. We observe that the very thin cosmologically allowed region of the traditional approach has been moved upwards (dark purple region) and gets broadened allowing for values of $m_0$ which are larger as compared to those of the traditional approach. The excluded region by $b \rightarrow s + \gamma$ data is shown as a hatched dark (cyan) region.
A relatively large region compatible with both Higgs and $g-2$ muon's data and also
$b \rightarrow s + \gamma$ data is allowed characterised by larger $m_0$ values as compared to the standard analysis. 
In figure \ref{fig4} we take $tan \beta = 55$ and we observe that the cosmologically allowed region according to the traditional scheme 
(light green stripe on the left) is now shifted upwards and to the right (dark purple region). Even from this rather small, but characteristic, sample it becomes apparent that conventional and Super-Critical String  Cosmology occupy rather different regions of the allowable supersymmetric parameter space, thus strengthening considerably the LHC potential for major discoveries.

As a last remark, in our treatment we 
find that, at temperatures $T_N \simeq 1 \; MeV$, radiation prevails over ordinary matter by almost seven orders of magnitude as demanded by Primordial Nucleosynthesis. It is worth noting that the radiation to matter ratio depends rather sensitively on the value of $w$ and it is remarkable that the cosmologically interesting values for $w$, according to the current astrophysical data, coincide with those for which the photon to matter ratio for successful  Primordial Nucleosynthesis is in the right ball park, while diluting at the same time the LSP relics by factors of  ${\cal{O}}(10)$.

\section{Conclusions}

In this paper we have shown that Supercritical String Cosmologies (SSC)
continue to provide a viable, dynamical framework to describe our wondrous Universe.
We have shown that for the set of parameters 
that provided the best fit to all supernovae
data until recently, we predict a rather smooth evolution of Dark Energy, for the last ten
billion years, thus in accordance with the very novel supernovae data~\cite{riessnew}. Furthermore,
we have shown that Superstring Cosmology off - equilibrium and time-dependent-dilaton effects~\cite{emnw,diamandis2}, with all parameters
fixed as explained above, dilute the neutralino Dark Matter, density to such a level
that while it relaxes the severe constraints imposed by conventional cosmology~\cite{susyconstr}, still
it keeps it in a SUSY parameter space exploitable by LHC. The fact 
that such a highly
non trivial dynamical framework as Supercritical String Cosmology, 
manages to provide such
a smooth varying Dark Energy, while fits all other available 
astrophysical data, including Primordial Nucleosynthesis, and at
the same time provides just the "right" dilution factor 
in order that neutralino Dark matter
continues to be the leading Dark Matter candidate~\cite{EHNOS,susyconstr}, 
and almost no dilution for baryons, is arguably remarkable. Clearly this observable entanglement in Supercritical String Cosmology involving Dark Energy and Supersymmetric Dark Matter may turn LHC to a smoking gun both for Supersymmetry and Supercritical Strings. 

\section*{Acknowledgements}

The work of A.B.L. and N.E.M.\ is
partially supported by funds made available by the European Social Fund (75\%)
and National (Greek) Resources (25\%) - EPEAEK~B - PYTHAGORAS and
by the
European Union through the Marie Curie Research and Training Network
\emph{UniverseNet} (MRTN-CT-2006-035863). That of D.V.N.\
is supported by D.O.E.\ grant DE-FG03-95-ER-40917.


\end{document}